\newcommand{\Msun}{\mathrm{M_{\odot}}}

\newcommand{\Mdot}{\dot{M}}
\newcommand{\simless}{\lesssim}
\newcommand{\simgreat}{\gtrsim}

\newcommand{\aap}{\textit{A\&A}}
\newcommand{\apj}{\textit{ApJ}}

\newcommand{\apjl}{\textit{ApJ}}

\newcommand{\mnras}{\textit{MNRAS}}
\newcommand{\nat}{\textit{Nature}}

\documentclass[dvips]{iaus}

\usepackage{graphicx}

\usepackage{url}
\usepackage{amssymb}
\usepackage{amsmath}
\usepackage{natbib}

\title[AGN Outflows] 
{Large-Scale Outflows from AGN: a link between central black holes 
and galaxies}

\author[Daniel Proga, Ryuichi Kurosawa \& Kentaro Nagamine]   
{Daniel Proga, Ryuichi Kurosawa\thanks{Present address:
    Department of Astronomy, Cornell University, Ithaca, NY
    14853-6801, USA.  email: {\tt kurosawa@astro.cornell.edu} } \and
  Kentaro Nagamine  }

\affiliation{Department of Physics and Astronomy, University of Nevada Las Vegas,
Box~454002, 4505~Maryland Pkwy, Las Vegas, NV 891541-4002, USA \\
email: {\tt {dproga, rk, kn}@physics.unlv.edu} }

\pubyear{2009}
\volume{267}  
\pagerange{119--126} 
\jname{Co-Evolution of Central Black Holes and Galaxies}
\editors{B.M.\ Peterson, R.S.\ Somerville, \& T.\ Storchi-Bergmann, eds.}

\begin{document}

\maketitle

\begin{abstract}
We summarize the results from numerical simulations of mass outflows
from AGN. We focus on simulations of outflows driven by radiation 
from large-scale inflows. We discuss the properties of
these outflows in the context of the so-called AGN feedback problem.
Our main conclusion is that this type of outflows 
are efficient in removing matter but inefficient in removing energy.


\end{abstract}

\firstsection 

\section{Introduction}
\label{sec:Introduction}

AGN produce the powerful outflows of the electromagnetic radiation that
in tunr can drive outflows of matter. These outflows and
the central location of AGN in their host galaxies 
imply that AGN can play a key role in determining the physical conditions in
the central region of the galaxy, the galaxy  as a whole, and 
even intergalactic scales
(e.g., \citealt{ciotti:1997}, \citealt{Ciotti:2009}; 
\citealt{Sazonov:2005}; \citealt{Springel:2005b};
\citealt{Fabian:2006b}; \citealt{Merloni:2008}, and references therein). 

Outflows from AGN are an important link between the central 
and outer parts of the galaxy because AGN
are powered by mass accretion onto a super massive black hole (SMBH).
Thus the SMBH 'knows'
about the galaxy through the accretion flow whereas
the galaxy knows about the SMBH through the outflows powered by accretion.
To quantify this connection, let us express
the radiation luminosity due to accretion as  
\begin{equation}
L_\mathrm{a}= \epsilon_{\mathrm{r}} c^2 \Mdot_{\mathrm{a}},
\label{eq:L-acc}
\end{equation}
where we invoke the simplest assumption such that the luminosity is
proportional to the mass accretion rate ($\Mdot_{\mathrm{a}}$) 
and a radiative (or the rest-mass conversion) efficiency
($\epsilon_{\mathrm{r}}$).

To estimate $\Mdot_\mathrm{a}$, one often adopts the analytic 
formula by \citet{Bondi:1952} who considered spherically symmetric 
accretion from a non-rotating polytropic gas with uniform density $\rho_\infty$
and sound speed $c_\infty$ at infinity. Under these assumptions, 
a steady state solution to the equations 
of mass and momentum conservation exists 
with a mass accretion rate of
\begin{equation}
\Mdot_\mathrm{B}= \lambda \, 4 \pi r^2_\mathrm{B} \rho_\infty c_\infty,
\label{eq:mdot_bondi}
\end{equation} 
where $\lambda$ is a dimensionless parameter
that, for the Newtonian potential, depends only on the adiabatic index,
$\gamma$ (cf.~\citealt{Bondi:1952}). 
The Bondi radius, $r_\mathrm{B}$, is defined as
\begin{equation}
r_\mathrm{B}=\frac{G M}{c^2_\infty}
\label{eq:r_bondi}
\end{equation}
where $G$ is the gravitational constant and $M$ is the mass of the accretor.

To quantify AGN feedback, one can measure its efficiency in changing
the flow of mass, momentum, and energy. 
The mass feedback efficiency $\epsilon_{\mathrm{m}}$ is
defined as the ratio of the mass-outflow rate at the outer boundary
$\dot{M}_{\mathrm{out}}$ to the mass-inflow
rate at the inner boundary $\dot{M}_{\mathrm{in}}$, i.e.,
\begin{equation}
  \epsilon_{\mathrm{m}}=
  \dot{M}_{\mathrm{out}}/\dot{M}_{\mathrm{in}}\,.
  \label{eq:eff-mass}
\end{equation}

Here, we consider only the energy and momentum carried out by matter.
Therefore, the momentum feedback efficiency 
($\epsilon_{\mathrm{p}}$) is defined
as the ratio of the total wind momentum $p_{\mathrm{w}}$ 
to the total radiation momentum ($L_{\mathrm{a}}/c$), i.e.,
\begin{equation}
  \epsilon_{\mathrm{p}}=p_{\mathrm{w}}/\left(L_{\mathrm{a}}/c\right)\,.
  \label{eq:eff-momentum}
\end{equation}

Finally, the total energy feedback 
efficiency $\epsilon_{\mathrm{t}}$ is defined
as the ratio between the sum of the kinetic power (kinetic energy flux) 
$P_{\mathrm{k}}$ and thermal energy power (thermal energy flux) 
$P_{\mathrm{th}}$ and the accretion luminosity of the system
$L_{\mathrm{a}}$  (Eq.~{[}\ref{eq:L-acc}]), i.e., 
\begin{equation}
  \epsilon_{\mathrm{t}}=\left(P_{\mathrm{k}}+P_{\mathrm{th}}\right)/L_{\mathrm{a}}\,,
  \label{eq:eff-total-energy}
\end{equation}
where
\begin{equation}
  \epsilon_{\mathrm{k}}=P_{\mathrm{k}}/L_{\mathrm{a}}
  \label{eq:eff-kinetic-energy}
\end{equation}
 and 
\begin{equation}
  \epsilon_{\mathrm{th}}=P_{\mathrm{th}}/L_{\mathrm{a}}\,,
  \label{eq:eff-thermal-energy}
\end{equation}
It follows that
$\epsilon_{\mathrm{t}}=\epsilon_{\mathrm{k}}+\epsilon_{\mathrm{th}}$.

Several very sophisticated simulations of feedback effects were recently 
performed, e.g., \citet{Springel:2005b} (SDH05 hereafter), 
\citet{DiMatteo:2005}, and \citet{Booth:2009} (BS09 hereafter).
These simulations follow merging galaxies in which many processes were 
included, for example, star formation, radiative cooling in a complex 
multi-phase medium, BH accretion and feedback. In addition, these local and 
global processes were connected. However, all of this was possible 
at the cost of crude phenomenological realizations of some processes
and the spatial resolution at a level larger than $r_\mathrm{B}$.
Consequently, the  efficiencies were assumed instead of 
being computed.

A main result of these simulations is that the famous $M_{\mathrm{BH}}-\sigma$ 
relation (e.g., \citealt{Ferrarese:2000}; \citealt{Gebhardt:2000}; 
\citealt{Tremaine:2002}) was reproduced. In addition, the BH mass 
is little affected by the details of star formation and supernova 
feedback.

This is a great success of the current cosmological simulations 
models. However, as pointed out above (see also others
e.g., \citealt{Begelman:2005})
the key feedback processes in the models represent ``subgrid'' physics. 
Therefore, 
in these cosmological and galaxy merger simulations 
AGN feedback cannot be directly related to AGN physics. 

On the other hand, models that aim to provide insights to AGN
physics do not include galaxy but rather focus on $r_\mathrm{B}$ or
even smaller scales (e.g., \citealt{Proga:2007b};
\citealt{Proga:2008}; \citealt{Kurosawa:2008};
\citealt{Kurosawa:2009, Kurosawa:2009b}). Thus they cannot be
directly related to AGN feedback on large scales. However, these
smaller scale simulations can be used directly to measure the feedback 
efficiencies and in turn
to quantify the effects that 
are assumed or parametrized in large scale simulations.

Here, we summarize the main findings from \citet{Kurosawa:2009c} where we
attempted to determine if AGN can supply energy in the form and amount 
required by the cosmological and galaxy merger simulations.
Our approach was to measure $\Mdot_{\mathrm{a}}$, and 
various feedback efficiencies based on direct
simulations of inflows and outflows in AGN, on sub-parsec and parsec
scales performed by \citet{Kurosawa:2009b} (KP09 hereafter). 

\section{AGN Models in Recent Cosmological Simulations}

Before we present our results, we briefly summarize what is 
required by the cosmological and galaxy merger simulations.

\label{sec:AGN-SPH}

\subsection{Mass Accretion Rates}

\label{sub:Mdot-SPH}

As we mentioned above,
cosmological and galaxy merger simulations (e.g., SDH05;
\citealt{Sijacki:2007}; BS09),
the actual physical process of the mass-accretion onto the BH is not
explicitly modeled because of a relatively poor resolution. 
These simulations rely on a separate analytical model to describe
the small scale processes. The unresolved accretion process
is usually described by a Bondi-Hoyle-Little formulation 
(e.g., \citealt{Bondi:1952}). 

One can illustrate some important issues related
to numerical realization of this  process by considering
a simpler Bondi accretion problem. 
Then, $\dot{M}_{\mathrm{a}}$ can be written as  
Eq.~(\ref{eq:mdot_bondi}). 
The dimensionless constant $\lambda$ (in Eq.~[\ref{eq:mdot_bondi}])
depends on $\gamma$ but is of order of unity. 
The Bondi formula relates $\dot{M}_{\mathrm{a}}$
of a BH located at
the center to the gas density and the sound speed (or equivalently the
temperature) of the gas at a large scale.

However in cosmological simulations,
the Bondi accretion is evaluated as 
\begin{equation}
  \dot{M_{\mathrm{S}}}=\alpha\,\frac{4\pi
    G^{2}M_{\mathrm{BH}}^{2}\rho}{c_{\mathrm{s}}^{3}}
\label{eq:mdot_bondi_sph}
\end{equation}
where $\rho$ and $c_{\mathrm{s}}$ are the density and the sound
speed estimated near the BH using the surrounding smoothed particle
hydrodynamics (SPH) gas
particles. Note that the expression contains {}``the dimensionless
parameter'' $\alpha$ which is different from $\lambda$ in
Eq.~(\ref{eq:mdot_bondi}). 
SDH05 introduced $\alpha$ parameter
to overcome the gap in the scale sizes between the numerical resolution
and the Bondi accretion regime. In a typical cosmological or galaxy
merger SPH simulation, the smoothing length ($\sim10^{3}$~pc) is much larger
than the gravitational radius of influence or the Bondi radius, 
$r_{\mathrm{B}}$, which is $\sim2$~pc.  
If we assume the gas located at a large distance,
heated by the AGN radiation, is {}``Comptonized''
($T\approx2\times10^{7}$~K), the corresponding speed of sound
(assuming $\gamma=5/3$) is 
relatively high ($\sim500\,\mathrm{km\, s^{-1}}$). 

It was found find that a very large factor of $\alpha$
is required for low-mass BHs to grow their masses; hence, the problem
is not strictly a Bondi accretion problem. Most of the AGN feedback
model in the cosmological simulations mentioned above assume a constant
value of $\alpha=100$ (see also Table~2 in BS09) except for BS09
who allow $\alpha$ to depend on the value of local gas density.
The assumption of a very large value of $\alpha$ becomes inadequate
when the local gas density is higher than that required by the formation
of the a cold interstellar gas phase, and when the cosmological simulation
does resolve the Jean length and the Bondi radius (BS09). 

BS09 abandoned the assumption of constant $\alpha$ in
Eq.~(\ref{eq:mdot_bondi_sph}), 
and introduced the following parametrization of $\alpha$. 
\begin{equation}
  \alpha=
  \begin{cases}
    1 & \mathrm{for\, n_{\mathrm{H}}<n_{\mathrm{H}}^{*}}\\
    \left(n_{\mathrm{H}}/\mathrm{n_{\mathrm{H}}^{*}}\right)^{\beta} &
    \mathrm{for}\, n_{\mathrm{H}}\geq n_{\mathrm{H}}^{*}
  \end{cases}
  \label{eq:alpha-booth}
\end{equation}
 where $n_{\mathrm{H}}$ and $n_{\mathrm{H}}^{*}$ are the number
density of hydrogen and the critical hydrogen number density above
which the gas is expected to become multi-phase, and star formation is
expected to begin via contraction of gas due to thermo-gravitational
instability (cf. \citealt{Schaye:2004}).
The critical density is chosen as $n_{\mathrm{H}}^{*}=0.1\,\mathrm{cm^{-3}}$,
 i.e., the corresponding critical hydrogen density is
 $\rho_{\mathrm{H}}^{*}=1.7\times10^{-25}\,\mathrm{g\, cm^{-3}}$. 
The best fit models of BS09 to some observations 
(e.g., the $M_{\mathrm{BH}}$--$\sigma$ relation) gives $\beta=2.0$. 
Note that the new parametrization of
$\alpha$ in Eq.~(\ref{eq:alpha-booth}) provides an additional
density dependency of the mass-accretion rate in 
Eq.~(\ref{eq:mdot_bondi_sph}).
In the formulation of BS09, $\Mdot_{\mathrm{a}}$
 steeply depends
on the density of the surrounding gas i.e., $\dot{M}\propto\rho^{3}$,
for $\rho>\rho_{\mathrm{H}}^{*}$ while the formulation of SDH05
and the original Bondi accretion model (Eq.~{[}\ref{eq:mdot_bondi}]) always
give a linear dependency i.e., $\dot{M}\propto\rho$. In most of the
AGN accretion models in the cosmological simulations (e.g., SDH05;
BS09), the highest $\Mdot_{\mathrm{a}}$ is limited to the Eddington
rate, i.e., 
\begin{equation}
  \dot{M}_{\mathrm{Edd}}=\frac{4\pi
    GM_{\mathrm{BH}}m_{\mathrm{p}}}{\epsilon_{\mathrm{r}}\sigma_{\mathrm{T}}c}
\label{eq:mdot-Eddington}
\end{equation}
 where $m_{\mathrm{p}}$, $\epsilon_{\mathrm{r}}$, $\sigma_{\mathrm{T}}$
and $c$ are the proton mass, the radiative efficiency (the rest mass
to radiation conversion efficiency), the Thomson cross-section and
the speed of light, respectively. The Eddington ratio ($\Gamma$) is
defined as  $\Mdot_{\mathrm{a}} / \dot{M}_{\mathrm{Edd}}$.

To illustrate how $\Mdot_{\mathrm{a}}$ depends on the density ($\rho$)
in different approaches,
Fig.~1 (left panel) compares results from 
the models by \citet{Bondi:1952}, SDH05 and BS09. 
The mass of the BH is assumed as
$M_{\mathrm{BH}}=10^{8}\,\Msun$. 
In all three models, the speed of sound $c_{\mathrm{s}}$ is set to
$520\,\mathrm{km\, s^{-2}}$, which corresponds to that of Comptonized
gas temperature $T\approx2\times10^{7}$~K with 
$\gamma=5/3$. In the modified Bondi accretion models of SDH05
and BS09, $\Mdot_{\mathrm{a}}$ is limited by the Eddington rate
(Eq.~{[}\ref{eq:mdot-Eddington}]), and the radiative efficiency
$\epsilon_{\mathrm{f}}$ is set to $0.1$. The dimensionless parameter
$\alpha=100$ is adopted for the model of SDH05, and $\beta=2.0$
is adopted in the model of BS09. The figure shows that the mass-accretion
rate of SDH05 is larger than the Bondi accretion rate for $\rho\simless10^{-20}\,\mathrm{g\, cm^{-3}}$.
On the other hand, $\Mdot_{\mathrm{a}}$ by BS09 is similar to
that of the Bondi accretion rate (off by a factor of $\lambda=1/4$
in Eq.~{[}\ref{eq:mdot_bondi}]) in the low-density regime ( $\rho\simless10^{-25}\,\mathrm{g\, cm^{-3}}$),
but it is significantly larger than the Bondi accretion rate for the
density range of $10^{-25}\simless\rho\simless10^{-20}\,\mathrm{g\, cm^{-3}}$.
The densities above which the accretion proceeds at the Eddington rate
are $\sim9.5\times10^{-23}\,\mathrm{g\, cm^{-3}}$ and $\sim6.3\times10^{-24}\,\mathrm{g\, cm^{-3}}$
for the models of SDH05 and BS09, respectively. However, these values
change depending on the adopted value of $c_{\mathrm{s}}$. In the
Bondi accretion model, $\Mdot_{\mathrm{a}}$ reaches the Eddington
rate at much higher density ($\rho\sim10^{-20}\,\mathrm{g\, cm^{-3}}$). 


\begin{figure}
\begin{center}

\includegraphics[clip,width=0.45\textwidth]{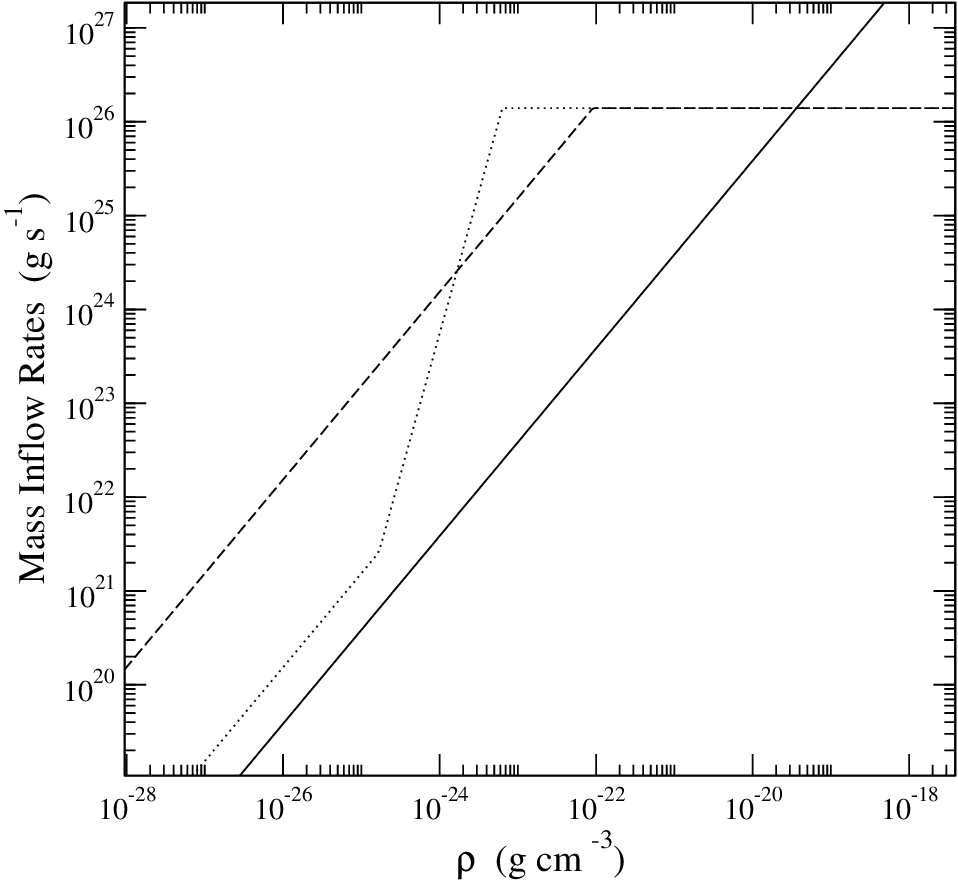}
\hspace{0.5cm}
\includegraphics[clip,width=0.45\textwidth]{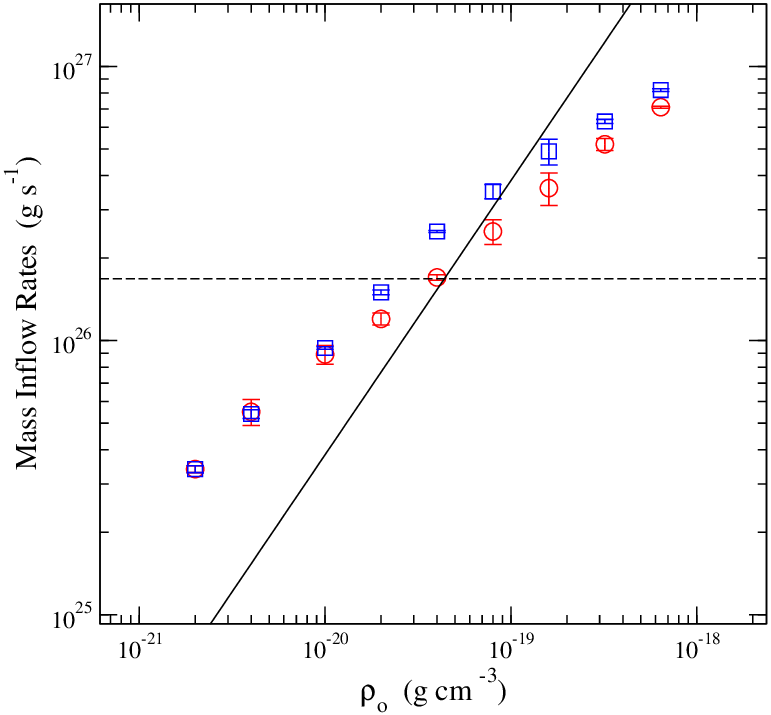}

\end{center}

\caption{{\it Left panel:} The mass-accretion rates from various models:
the Bondi accretion model (\citealt{Bondi:1952}) (\emph{solid}), 
the modified Bondi accretion model by SDH05 (\emph{dashed}) 
and that by BS09 (\emph{dotted}), as described in 
Eqs.~(\ref{eq:mdot_bondi}), (\ref{eq:mdot_bondi_sph})
and (\ref{eq:alpha-booth}), respectively. The mass-accretion rates
are computed as a function of $\rho$ at a radius $r_{\mathrm{o}}$
which is much larger than $r_{\mathrm{B}}$, i.e.,
$r_{\mathrm{o}}\gg r_{\mathrm{B}}$. In all models, the speed of sound
$c_{\mathrm{s}}$ at $r_{\mathrm{o}}$ is set to $520\,\mathrm{km\, s^{-2}}$,
The mass of the BH
is set to $M_{\mathrm{BH}}=10^{8}\,\Msun$. While $\alpha=100$ is
adopted in the model of SDH05, $\beta=2.0$ is adopted
in the model of BS09. In the modified Bondi accretion
models of SDH05 and BS09, the mass-accretion
rates are limited to the Eddington rate
(Eq.~{[}\ref{eq:mdot-Eddington}]). 
{\it Right panel:} 
Comparison of the mass-inflow rates found in the HD simulations 
with those predicted by the Bondi accretion model (\citealt{Bondi:1952})
(\emph{solid line}) and with those adopted by SDH05
and BS09 (\emph{dashed line}). In the density range
of the models considered here, $\Mdot_{\mathrm{a}}$ 
adopted by SDH05 and BS09 are limited by the
Eddington rate (Eq.~{[}\ref{eq:mdot-Eddington}]); hence, the line
is flat. The mass-inflow rates
at the inner boundary (\emph{circles}) and those at the outer boundary
(\emph{squares}) of the computational domain are shown as a function
of the outer boundary density $\rho_{\mathrm{o}}$. 
The simulations predict $\Mdot_{\mathrm{a}}$ that are 
very similar to the Bondi accretion rates,
but the models have a less steeper dependency on the density. The
Bondi mass-accretions rates and those of SDH05 and
BS09 are computed for the gas with the Comptonized temperature
$T=2\times10^{7}$~K and with $\gamma=5/3$. 
The figure is from \citet{Kurosawa:2009c}.
}

\end{figure}


\subsection{AGN Feedback Models}

\label{sub:AGN-FB-SPH}

Once $\Mdot_{\mathrm{a}}$ is evaluated, one  can estimate
the amount of the accretion power that is released and deposited.
In SDH05 and BS09 (also in many others, cf.~Table~2 in BS09),  
$\epsilon_{\mathrm{r}}=0.1$
(e.g., \citealt{shakura:1973}; see also \citealt{Soltan:1982})
is assumed, and kept constant. A higher value of $\epsilon_{\mathrm{r}}$
($\sim0.2$) can be achieved in an accretion model with a thin disk
and a rapidly rotating BH (e.g., \citealt{Thorne:1974}).
Recent observations suggest a wide range of $\epsilon_{\mathrm{r}}$:
$0.07$ (\citealt{Martinez-Sansigre:2009}), $0.30$--$0.35$
(\citealt{Wang:2006}).
On the other hand, \citet{Cao:2008} find $\epsilon_{\mathrm{r}}$
is relatively low ($\sim0.08$) for $M_{\mathrm{BH}}<10^{8}\,\Msun$
and relatively high 
($\simgreat0.18$) for $M_{\mathrm{BH}}\simgreat10^{9}\,\Msun$.
The exact mechanism coupling the accretion luminosity of a BH
and the surrounding gas is not well known. Therefore,
SDH05 and BS09 simply assumed that $L_{\mathrm{a}}$ couples only
thermally (and isotropically) to the surrounding. Using
Eq.~(\ref{eq:L-acc}), the rate of energy deposition to the
surrounding (the AGN feedback rate) in SDH05 is 
\begin{equation}
  \dot{E}_{\mathrm{f}}=\epsilon_{\mathrm{f}}\,
  L_{\mathrm{a}}=\epsilon_{\mathrm{f}}\epsilon_{\mathrm{r}}\dot{M}_{\mathrm{BH}}c^{2}
\label{eq:Edot-Feedback}
\end{equation}
 where $\epsilon_{\mathrm{f}}$ is the efficiency of the AGN energy
deposition to the surrounding gas, and is a free parameter which is to
be constrained by observations. BS09 find the models
with $\epsilon_{\mathrm{f}}=0.15$ matches observations (e.g., the Magorrian
relation and the $M_{\mathrm{BH}}-\sigma$ relations) very well, and
similarly SDH05 find $\epsilon_{\mathrm{f}}=0.05$ matches
observations well. 

\section{Results from Our Numerical Simulations}
\label{sec:Results}

In \citet{Kurosawa:2009c}, we 
used our physical two-dimensional (2-D) and time-dependent
hydrodynamical (HD) simulations of AGN flows to investigate the dependency
of the BH mass-accretion rate on the surrounding gas density, and to
measure the AGN feedback efficiencies in converting the accretion
luminosity into the outward fluxes of energy, momentum and mass. 
To do this, 
we simply analyzed the simulations  previously published in KP09.  
We note that 2-D simulations are consistent in many respects
with their three-dimensional (3-D) counterparts \citep{Kurosawa:2009}.

\subsection{Mass-Accretion Rates}

\label{sub:result-mdot}

The mass-inflow rates at the inner boundary 
$\dot{M}_{\mathrm{in}}\left(r_{\mathrm{i}}\right)$
and those at outer boundary $\dot{M}_{\mathrm{in}}\left(r_{\mathrm{o}}\right)$
from the HD simulations are plotted as a function of the outer boundary
density $\rho_{\mathrm{o}}$ in the right panel of Fig.~1. 
For a given value of $\rho_{\mathrm{o}}$,
$\dot{M}_{\mathrm{in}}\left(r_{\mathrm{i}}\right)$
and $\dot{M}_{\mathrm{in}}\left(r_{\mathrm{o}}\right)$ are not equal
to each other, but rather
$\dot{M}_{\mathrm{in}}\left(r_{\mathrm{i}}\right)<\dot{M}_{\mathrm{in}}\left(r_{\mathrm{o}}\right)$
because of an outflow. The lowest density model is an exception
since no outflow is formed in this model. For the higher density models,
an outflow forms, and not all the material entering from the outer
boundary reaches the inner boundary. A fraction of gas experiences
a strong radiation pressure and radiative heating, and the direction
of flow changes, forming an outflow.

The figure also shows $\Mdot_{\mathrm{a}}$ predicted by the
Bondi accretion model (Eq.~{[}\ref{eq:mdot_bondi}]) and those computed
from the formulations of SDH05 and BS09 (Eqs.~{[}\ref{eq:mdot_bondi_sph}]
and {[}\ref{eq:alpha-booth}]). The outer radius
is much smaller than that of a typical smoothing scale on a SPH cosmological
simulation ($\sim10^{3}$~pc), and the outer density values used
in our simulations are much larger than a typical local density at
BH in the SPH simulations. The higher density
at a 10~pc scale is required to produce an outflow.
For example, $\rho_{\mathrm{o}}$
must be greater than $2\times10^{-21}\,\mathrm{g\, cm^{-3}}$, 
to form an outflow with our system setup. In the density
range of the models considered here, $\Mdot_{\mathrm{a}}$ adopted
by SDH05 and BS09 are limited by the Eddington rate (Eq.~{[}\ref{eq:mdot-Eddington}]);
hence, the line is flat (cf.~left panel of Fig.~1).

We do not expect our solutions to reproduce 
the Bondi density dependency of the mass-inflow rate 
because they include the effects of radiative heating and radiation force.
However, the right panel of Fig.~1
shows the mass-inflow rates from our models
are very similar to those of the Bondi rates, i.e., the rates are
of the same order of magnitude. Interestingly our 
$\dot{M}_{\mathrm{in}}\left(r_{\mathrm{i}}\right)$
and the Bondi mass-accretion rate matches around 
$\rho_{\mathrm{o}}=4\times10^{-20}\,\mathrm{g\, cm^{-3}}$
which coincidentally corresponds to $\Gamma\approx1$. The accretion
rates from SDH05 and BS09 are Eddington (the rates corresponding
to $\Gamma=1$) in this density range. Therefore 
their corresponding lines also cross at the same point.

Fig.~1 (right panel) shows that our models have a weaker dependency of the
mass-inflow rates on the density than that of the Bondi accretion. 
The power-law fits of data points give the slope 
$q=0.52\left(\pm0.01\right)$ for $\dot{M}_{\mathrm{in}}\left(r_{\mathrm{i}}\right)$ and 
$q=0.56\left(\pm0.02\right)$ for $\dot{M}_{\mathrm{in}}\left(r_{\mathrm{o}}\right)$,
which are indeed much smaller than that of the Bondi accretion model,
i.e., $q=1$ (cf.~Eq.~{[}\ref{eq:mdot_bondi}]).


\subsection{Feedback Efficiencies}

\label{sub:efficiency-HD}

Let us now consider AGN feedback efficiencies in energy, momentum and
mass using the simulation results, as defined in
Eqs.~(\ref{eq:eff-mass})--(\ref{eq:eff-total-energy}). The
models used here show some degree of variability (typically $\sim10\%$
level). Therefore, the physical quantities used to compute the feedback
efficiencies are based on the time averaged values.

\subsubsection{Energy Feedback Efficiency}

\label{subsub:eff_energy}

Fig.~\ref{fig:eff_energy} (left panel) shows the energy feedback efficiencies,
$\epsilon_{\mathrm{t}}$, $\epsilon_{\mathrm{k}}$ 
and
$\epsilon_{\mathrm{th}}$ computed based on our models, as a function of the Eddington ratio
($\Gamma$).  For systems with relatively low Eddington
ratio ($\Gamma$$\simless0.4$), 
$\epsilon_{\mathrm{th}}>\epsilon_{\mathrm{k}}$.  On
the other hand, for systems with relatively high Eddington ratio
($\Gamma$$\simgreat0.6$), the kinetic feedback dominates the thermal
feedback by a factor of $\sim10$ to $\sim100$. The model with
$\Gamma=0.2$ does not form an outflow, indicating an approximate
$\Gamma$ value below which no outflow forms (with our system setup).

The energy feedback efficiencies increase as $\Gamma$ increases, but
the efficiencies saturate for $\Gamma\simgreat1$. The total energy
feedback efficiency peaks at $\Gamma\approx1$ with
$\epsilon_{\mathrm{t}}\sim10^{-4}$.  The flattening of the
efficiencies for $\Gamma\simgreat1$ is caused by the transition of the
inflow-outflow morphology to a {}``disk wind'' like configuration
for the higher $\Gamma$ models (see fig.~4 in KP09). Because of the mismatch between the
direction in which most of the radiation escapes (in polar direction)
and the direction in which the most of the accretion 
occurs in the system (the equatorial direction), the radiatively
driven outflows in the disk-wind-like configuration cannot increase
the outflow efficiency by increasing the accretion luminosity or
equivalently $\Gamma$. A similar behavior is found in the
$\dot{M}_{\mathrm{out}}\left(r_{\mathrm{o}}\right)$-- $\Gamma$ relation
of KP09.

\subsubsection{Mass Feedback Efficiency}

\label{subsub:eff_mass}

\begin{figure}
\begin{center}

\includegraphics[clip,width=0.45\textwidth]{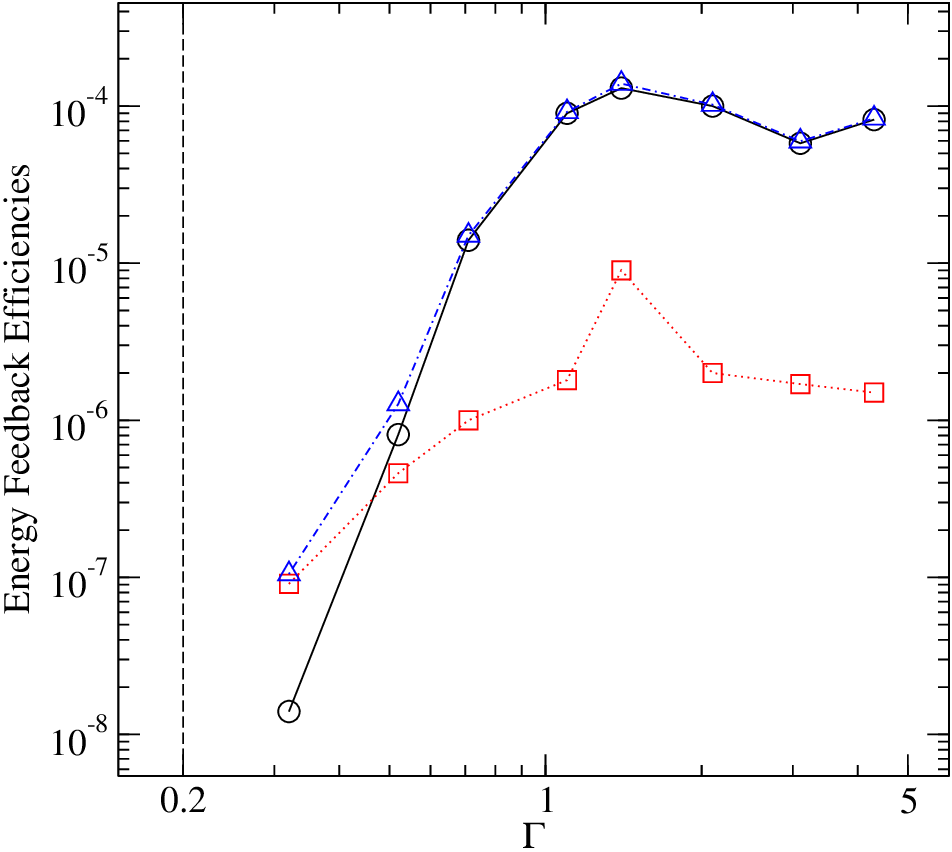}
\hspace{0.5cm}
\includegraphics[clip,width=0.45\textwidth]{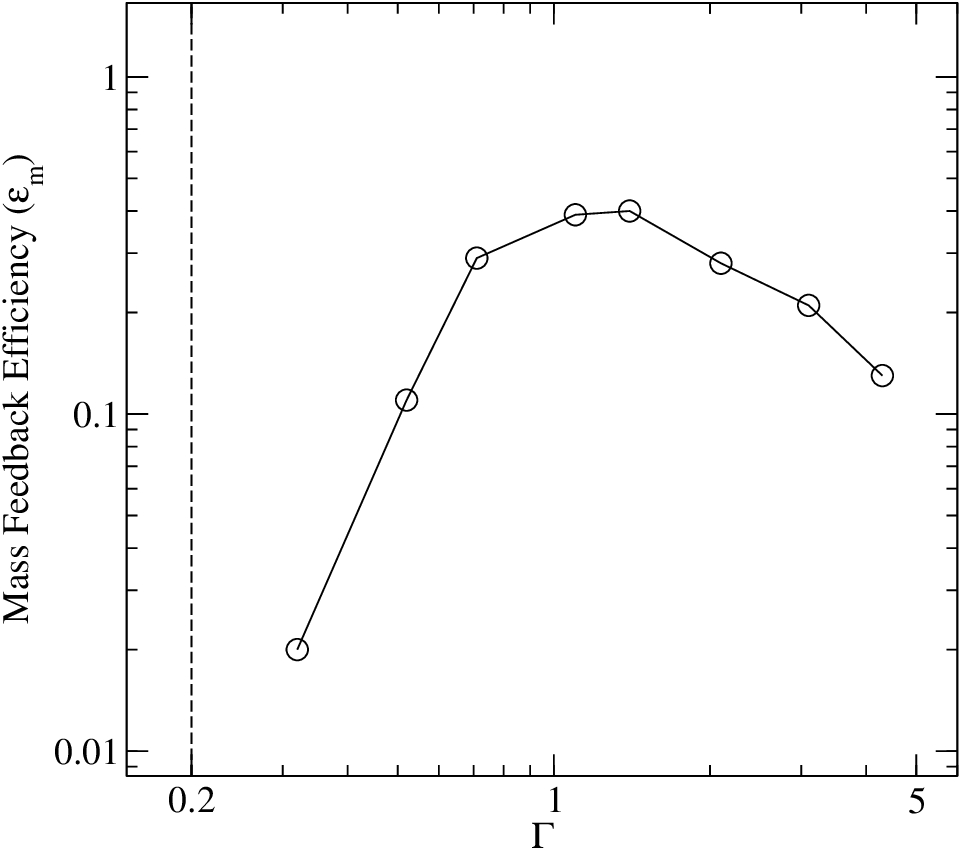}

\end{center}

\caption{\emph{Left panel:} the efficiencies of converting the BH
  accretion luminosity $L_{\mathrm{a}}$ to the rate of energy
  deposition to the surrounding gas are plotted as a function of the
  Eddington ratio ($\Gamma$). The panel shows the kinetic energy
  feedback efficiency $\epsilon_{\mathrm{k}}$ (\emph{circles}), the
  thermal energy feedback efficiency $\epsilon_{\mathrm{th}}$
  (\emph{squares}) and the total energy feedback efficiency
  $\epsilon_{\mathrm{t}}=\epsilon_{\mathrm{k}}+\epsilon_{\mathrm{th}}$
  (\emph{triangles}), separately (see
  Eqs.~{[}\ref{eq:eff-total-energy}], {[}\ref{eq:eff-kinetic-energy}],
  and {[}\ref{eq:eff-thermal-energy}]).  The maximum total energy
  feedback efficiency is $\sim10^{-4}$.  For the models with
  relatively low Eddington ratio ($\Gamma$$\simless0.4$), the thermal
  feedback is more efficient than the kinetic feedback
  ($\epsilon_{\mathrm{th}}>\epsilon_{\mathrm{k}}$).  For the models
  with relatively high Eddington ratio ($\Gamma \simgreat0.6$), the
  kinetic feedback is more efficient than the thermal feedback by a
  factor of $\sim10$ to $\sim100$. The model with $\Gamma=0.2$ does
  not form an outflow, and the vertical line (\emph{dashed}) at
  $\Gamma=0.2$ indicates an approximate $\Gamma$ value below which no
  outflow forms.  The flattening of the efficiencies at
  $\Gamma\approx1$ is caused by the transition of the inflow-outflow
  morphology to a {}``disk wind'' like configuration for the larger
  $\Gamma$ models (see fig.~4 in KP09).  
  \emph{Right panel}: the mass feedback efficiency
  ($\epsilon_{\mathrm{m}}$) plotted as a function the Eddington ratio
  ($\Gamma$). 
  The efficiency peaks around $\Gamma=1$ with the maximum efficiency
  value $\sim0.4$, i.e., $40\%$ of the total inflowing mass is
  converted to the outflows. The turn-around of
  $\epsilon_{\mathrm{m}}$ values is caused by the transition of the
  outflow morphology to a {}``disk wind'' like configuration for the
  larger $\Gamma$ models (cf. KP09). The model with
  $\Gamma=0.2$ does not form an outflow, and the vertical line
  (\emph{dashed}) at $\Gamma=0.2$ indicates an approximate $\Gamma$
  value below which no outflow forms.
  The figures are from \citet{Kurosawa:2009c}.
}

\label{fig:eff_energy}
\label{fig:eff_mass}
\end{figure}

Figure~\ref{fig:eff_mass} (right panel) shows $\epsilon_{\mathrm{m}}$
plotted as a function $\Gamma$. The dependency
of $\epsilon_{\mathrm{m}}$ on $\Gamma$ is very similar to that of
the momentum feedback efficiency (not shown here).
For $\Gamma\simless1$, the mass feedback efficiency $\epsilon_{\mathrm{m}}$
increases as $\Gamma$ increases, but it starts to decreases slightly
beyond $\Gamma\approx1$. The efficiency peaks around $\Gamma=1$
with the maximum efficiency value $\sim0.4$. In other words, about
$40\%$ of the total inflowing mass is redirected
to an outflow. The turn-around of $\epsilon_{\mathrm{m}}$ values is
caused by the transition of the outflow morphology to a disk-wind-like
configuration (see fig. 4 in KP09) for the larger $\Gamma$
models, as in the cases for the energy and momentum feedback
efficiencies (Sec.~\ref{subsub:eff_energy}).

The accretion process in our model is fundamentally different from
that of the Bondi accretion model and those adopted in the cosmological
simulations (e.g., SDH05; BS09) because in our model
an outflow and an inflow can simultaneously be formed while the Bondi
accretion model can only form an inflow. The model of SDH05 and others
can form either an inflow or an accretion (but not both simultaneously).

\section{Conclusions}

We  performed over thirty
number of axisymmetric, time-dependent radiation-hydrodynamical simulations of
outflows from a slowly rotating (sub-Keplerian) infalling gas driven 
by the energy and pressure of the radiation emitted by the AGN
(KP09). These simulations follow dynamics of gas under 
the influence of the gravity of the
central $10^8~\Msun$ black hole on scales from $\sim0.01$ 
to $\sim 10$~pc. They self-consistently couple the accretion-luminosity with
the mass inflow rate at the smallest radius (a proxy for 
$\dot{M}_{\mathrm{a}}$). 
A key feature of
the simulations is that the radiation field and consequently the gas
dynamics are axisymmetric, but not spherically symmetric.  Therefore,
the gas inflow and outflow can occur at the same time. 

The relatively large number of simulations allows us to investigate 
how the results depend on the gas density at the outer radius, 
$\rho_{\mathrm{o}}$.  
In a follow-up paper, \citet{Kurosawa:2009c}, we
measure and analyze the energy, momentum, and mass feedback
efficiencies of the outflows studied in KP09.
We compared our
$\dot{M}_{\mathrm{a}}$-$\rho_{\mathrm{o}}$ relation with that
predicted by the Bondi accretion model.  For high luminosities
comparable to the Eddington limit, the power-law fit
($\dot{M}_{\mathrm{a}} \propto \rho_{\mathrm{o}}^{q}$) to our models
yields $q\approx 0.5$ instead of $q=1.0$ which is predicted by the
Bondi model.  This difference is caused by the outflows which are
important for the overall mass budget at high luminosities.  The
maximum momentum and mass feedback efficiencies found in our models
are $\sim 10^{-2}$ and $\sim 10^{-1}$, respectively.  However, the
outflows are much less important energetically: the thermal and
kinetic powers in units of the radiative luminosity are $\sim 10^{-5}$
and $\sim 10^{-4}$, respectively. In addition, the efficiencies do not
increase monotonically with the accretion luminosity, but rather peak
around the Eddington limit beyond which a steady state disk-wind-like
solution exists.  Our energy feedback efficiencies are significantly
lower than 0.05, which is required in some cosmological and galaxy
merger simulations.  The low feedback efficiencies found in our simulations 
could
have significant implications on the mass growth of super massive
black holes in the early universe.

\end{document}